\documentclass[twocolumn,prl,showpacs,floatfix]{revtex4}
\usepackage{graphicx}
\usepackage{epsfig}
\usepackage{rotating}
\usepackage{amsmath}
\usepackage{amsfonts}
\usepackage{amssymb}
\usepackage{enumerate}
\usepackage{longtable}
\usepackage{dcolumn}% Align table columns on decimal point
\usepackage{bm}

\begin{document}

\title{On the entanglement across a cubic interface in $3+1$ dimensions}

\author{Trithep Devakul$^1$ and Rajiv R. P. Singh$^2$}
\affiliation{
    $^1$Department of Physics, Northeastern University, MA 02115, USA \\
$^2$Department of Physics, University of California Davis, CA 95616, USA}

\date{\rm\today}

\begin{abstract}
We calculate the area, edge and corner Renyi entanglement entropies in the ground state of the transverse-field
Ising model, on a simple-cubic lattice, by high-field and low-field series expansions. 
We find that while the
area term is positive and the line term is negative as required by strong subadditivity, the corner contributions
are positive in 3-dimensions. 
Analysis of the series suggests that the expansions converge up to the physical critical point 
from both sides.
The leading area-law Renyi entropies match nicely from the high and low field expansions at the critical 
point, forming a sharp cusp there. We calculate the coefficients of the logarithmic divergence associated with the corner entropy
and compare them with conformal field theory results with smooth interfaces and find a striking correspondence.
\end{abstract}

%\pacs{74.70.-b,75.10.Jm,75.40.Gb,75.30.Ds}

\maketitle

Interest in studies of many-body quantum-entanglement encompasses many areas of physics. On the one hand, these are important in
studies of black hole thermodynamics and the holographic principle \cite{ryu}. On the other, they are informative about topological
and quantum-critical phases of matter \cite{cardy,rmp-review,max}. 
They can provide measures of thermalization in systems far from equilibrium \cite{peschel} and form the
basis for classifying stability of quantum field theories under renormalization \cite{cardy-rev,grover14}. 
Yet, many questions remain, especially in
relating continuum field theory results on smooth manifolds with lattice-defined models with sharp edges and corners.

In the ground state of a gapped non-topological phase of a lattice statistical model,
bipartite entanglement entropy is entirely associated
with the boundaries between subsystems.\cite{wolf08} As we increase the dimensionality (D), various boundaries of
reduced co-dimension can arise. In case of 3-dimensional simple cubic lattice, the boundaries 
can consist of planar-surfaces, line-edges and point-corners (See Fig.~1). As the critical coupling is approached the entanglement 
entropies are expected to develop singularities.\cite{rmp-review,max} These singularities are weakest for the manifold with smallest 
co-dimensionality and they are most singular for corners or points in the boundary, in which case they are expected to diverge 
logarithmically, irrespective of dimensionality. An interesting question is: Are these corner logarithms related to logarithmic 
singularities in 3+1 continuum quantum-field theories with smooth interfaces?

Singularities in the entanglement entropy are extensively studied in 1D quantum models thanks
to conformal field-theory methods and density matrix renormalization group (DMRG).\cite{cardy,peschel} There have also been
several computational studies of the singularities in 2D,\cite{melko,series-prl,oitmaa,nlc,nlc-dmrg,roscilde,wessel,devakul,inglis} 
and strong support for
universal behavior has emerged. 
Most of the studies so far
have been for Renyi rather than Von Neumann entanglement entropy as the former are expected to be 
just as good at capturing universal behavior. Recent work by Chandran et al \cite{chandran} has raised the possibility
of a fictitious Renyi-index dependent transition, which could spoil their universality.
The numerical evidence presented here suggests that the transition point is independent of the Renyi index and coincides with the physical
critical point \cite{footnote1}. 

\begin{figure}
\begin{center}
\includegraphics[width=7cm]{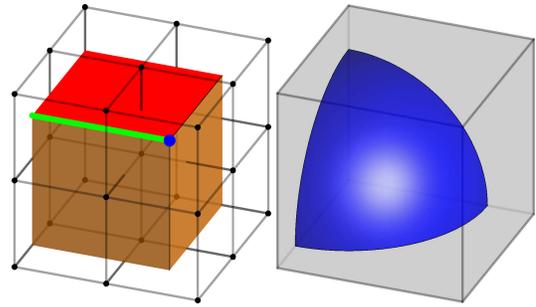}
\caption{\label{fig0}
Interfaces on a simple cubic lattice are made up of planes (shaded region), edges (solid line) and corners (circle). In a continuum
a sphere represents a smooth interface. Are log singularities for a cubic interface on a lattice related
to those for a spherical interface in the continuum?
}
\end{center}
\end{figure}

We study Renyi entanglement entropies of the transverse-field Ising model on the
simple-cubic lattice, by high and low-field series expansions.\cite{book,series-reviews}
These methods enable separate calculations of area, edge and corner entropies, thus allowing us
to isolate weaker terms associated with line-edges and point-corners.

The leading area-law entropy must be positive. By simply generalizing the strong subadditivity argument 
for corners in the 2-dimensional case,\cite{hastings} one
can show that the line entropy must be negative in three-dimensions. Although, the strong subadditivity is
only proven for the von Neumann entropy, the Renyi entropies also obey the predicted signs. We find that the entropies also
alternate in sign as we go from area to edge to corner entropy. 

Our series analysis results are 
in good agreement with the leading `area-law' term having a $1/\xi^2$ singularity, as expected from scaling arguments.\cite{series-prl}
For the all important corner term, we simply bias the series to have the expected
logarithmic singularity at the critical point and estimate the coefficient of log divergence. 

From the point of view of critical phenomena, $3+1$-D is at the boundary where
Gaussian or free fixed point becomes stable.\cite{fisher} Thus, a correspondence with free field
theory is expected, modulo logarithmic corrections. One should note that even in $2+1$-D, the log coefficients
for free field theory and Ising criticality are not far from each other \cite{nlc}. However, we are not aware
of a field theory calculation for a lattice model with a corner. Conformal field theory has
been used to study logarithmic singularity associated with smooth interfaces\cite{tarun,cft1,huerta}.
On a lattice, we can define a closed surface consisting
of a cube, with $8$ corners. Quite remarkably, we find that the log singularity in the corner entropy for a cube, and 
its dependence on the Renyi index, is quite close to $1/8$th of
the result for a sphere. This is potentially a deep result and deserves further theoretical attention\cite{tarun}.

%%%%%%%%%%%%%%%%%%%%%%%%
%Definitions for Area, Line and Corner entropies

\begin{figure}
\begin{center}
\includegraphics[width=7cm,angle=270]{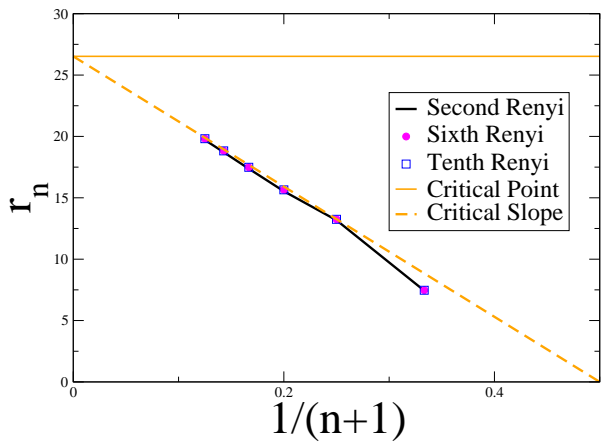}
\caption{\label{fig1} 
Ratio of successive series coefficients in the variable $\lambda^2=(J/h)^2$ for the `area-law' coefficient of various Renyi entropies.
Only 2nd, sixth and tenth Renyi entropies are shown. The difference between successive Renyi entropies is very tiny.
The critical-point value, known from the study of other ground state properties, is shown by a solid line. 
Assuming standard scaling for the singular part of the `area-law' term, the ratio of coefficients must asymptotically approach the 
critical-point with the slope given by the dashed line.
}
\end{center}
\end{figure}

\begin{figure}
\begin{center}
\includegraphics[width=7cm,angle=270]{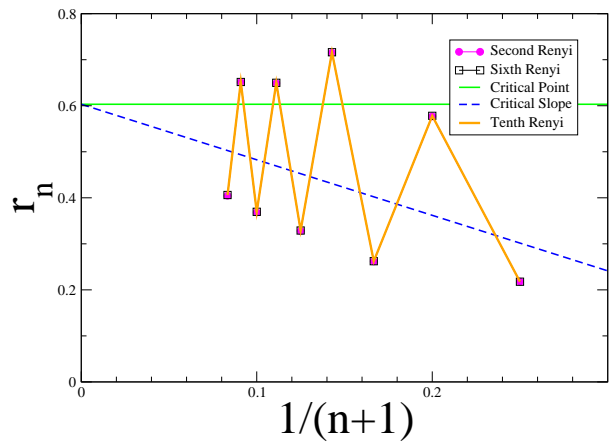}
\caption{\label{fig2} 
Ratio of successive series coefficients in the variable $x^2=(h/4J)^2$ for the `area-law' coefficient of various Renyi entropies.
Only second, sixth and tenth Renyi entropies are shown. The difference between them is very tiny.
The critical-point value, known from the study of other ground state properties, is shown by a solid line. 
Assuming standard scaling for the singular part of the `area-law' term, the ratio of coefficients must asymptotically approach the 
critical-point with the slope given by the dashed line.
}
\end{center}
\end{figure}

The transverse-field Ising Hamiltonian is
\begin{eqnarray}
{\cal H}&=& -h\sum_i S_i^x -J\sum_{\langle i,j\rangle} S_i^z S_j^z .
\end{eqnarray}
We expand around both the $J=0$ and the $h=0$ limits of the model.
Various ground state properties can be calculated as a power series 
expansion in the variables $\lambda=J/h$ or $x=h/4J$\cite{book,series-reviews,oitmaa}.
When a system is bipartitioned into two subsystems $A$ and $B$,
the $\alpha$th Renyi entropy is defined as
\begin{eqnarray}
    S_\alpha(A) &=& \frac{1}{1-\alpha}\ln \text{Tr}\left( \hat{\rho}^\alpha_A \right),
\end{eqnarray}
where $\hat{\rho}_A = \text{Tr}_B \left|{\Psi}\right\rangle \left\langle{\Psi}\right|$ is the reduced density matrix for subsystem $A$.

When the infinite system is bipartitioned by a plane (such as the XY plane), the ground state has an entropy per unit area.
To define an entropy per unit length, we consider dividing the system by two perpendicular planes. The area contributions
can be canceled out analogous to the calculation of corner entropy in 2D\cite{oitmaa,nlc-dmrg,devakul}
leaving one with an edge-entropy associated with a $90$ degree edge.
To define a $\pi/2$ solid angle corner entropy, we consider the intersection of three perpendicular planes. Once again,
the area and edge entropies can be canceled out by suitable subtraction, leaving one with the
corner entropy. If one has a closed cubic interface with large linear dimension, it can be decomposed into planar surfaces, 
$90$ degree edges and $\pi/2$ corners. Within perturbation theory, the total entanglement across the cube is the
sum of each contribution.

Let the entropy per unit boundary-area be defined as $a_\alpha=S_\alpha/A$, the entropy per unit length
for a $90$ degree edge be defined as $s_\alpha=S_\alpha/L$, and the corner entropy for a $\pi/2$ solid-angle 
be defined as $c_\alpha$.
%The series coefficients are defined as,
%\begin{equation}
%a_\alpha=\sum_n a(\alpha,n)\lambda^{2n},
%\end{equation}
%\begin{equation}
%s_\alpha=\sum_n b(\alpha,n)\lambda^{2n},
%\end{equation}
%and
%\begin{equation}
%c_\alpha=\sum_n c(\alpha,n)\lambda^{2n}.
%\end{equation}
%Non-zero coefficients to order $\lambda^{14}$ for $\alpha=2$ and $3$ are given in tables 1-3. 
Series expansion coefficients for $a_\alpha$, $s_\alpha$ and $c_\alpha$ are calculated complete
to order $\lambda^{14}$ in the high-field expansions and to order $x^{22}$ in the low-field
expansions.

%%%%%%%%%%%%%%%%%%%%%%

\begin{figure}
\begin{center}
\includegraphics[width=7cm,angle=270]{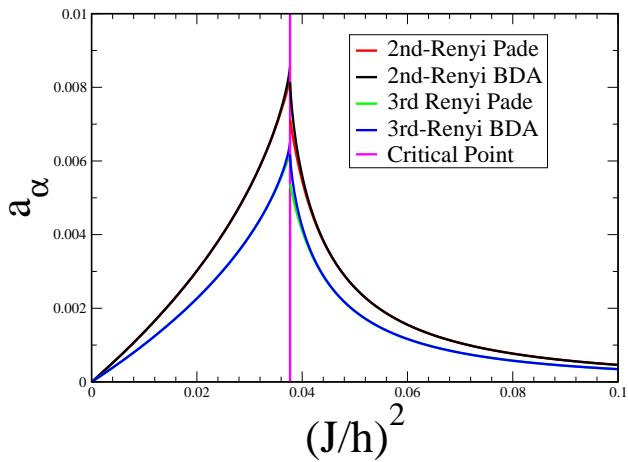}
\caption{\label{fig3} 
Plots of the `Area-law' or entanglement entropy per unit boundary area
for the second and third Renyi indices. Results of both small field and large field expansions
are shown. The vertical solid line shows the known critical value. A Pade approximant
and a biased differential approximant (BDA) are shown in each case for comparison. One can see that the critical singularities only
show up very close to the critical point. 
}
\end{center}
\end{figure}

\begin{figure}
\begin{center}
\includegraphics[width=7cm,angle=270]{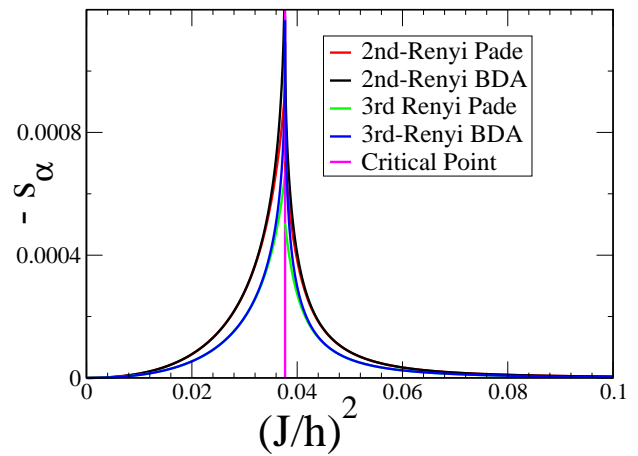}
\caption{\label{fig4}
Plots of the `edge' entanglement entropy or entanglement entropy per unit length at an edge in the interface.
Results of both small field and high field expansions
are shown. The vertical solid line denotes the known critical value. A Pade approximant
and a biased differential approximant (BDA) are shown in each case for comparison.
}
\end{center}
\end{figure}

We first study the critical point of the leading entropy, that is the area-law term, to find where it becomes singular.
To locate this critical point, we calculate the ratio of coefficients $r_n=a_{n}/a_{n-1}$. These
coefficients should approach $1/\lambda_c^2$ or $1/x_c^2$ as $n\to\infty$. The critical point of the model is well known
from previous series expansion study\cite{oitmaa-sc}
to be at $1/\lambda_c=5.15$. In Fig.~2 and Fig.~3, we show the ratio of coefficients
from the high and low field expansions respectively. The value of the known critical coupling is
also shown. Also shown as a dashed-line is the asymptotic slope along which the $r_n$ should approach the critical value
as $n\to\infty$, assuming a $1/\xi^2$ singularity with $\nu=1/2$.\cite{fisher,zinn-justin} From the high-field expansion,
we find that the numerical values are in very
good agreement with the expectations from scaling theory. Note that the data include ratios for second, sixth and tenth Renyi entropies.
Evidently there is no discernible dependence of the critical point on the Renyi index. However, we should note that
a very small variation in the critical point would probably not be distinguishable in our study.

The low-field expansions show a strong alternation, which is not unusual for series expansions around an ordered phase
(also sometimes called a low temperature expansion\cite{book}).
However, they also provide strong support for the idea that the singularities happen at the critical point and there
is no variation in the critical point with the Renyi index.

The area and line coefficients are analyzed both by simple Pade approximants that lack critical behavior and a biased
differential approximant, which builds in a power-law singularity at the critical point. They are shown in Fig.~4
and Fig.~5 respectively. 
Outside the critical region, where simple Pade approximants and biased differential approximants agree,
our results should be highly accurate. 
One finds that the critical behavior only sets in when one is within a few percent of the critical point. 
Furthermore, we find that the leading `area-law' entropies nicely meet together from the high and low temperature sides, 
forming a sharp cusp there.
This is further evidence that the only singularity in the system is at the physical critical point.
We do not find any evidence for a log singularity in the line term,
although, it would be difficult for series
analysis to reliably deal with  coexisting power-law and logarithmic singularities.

\begin{figure}
\begin{center}
\includegraphics[width=7cm,angle=270]{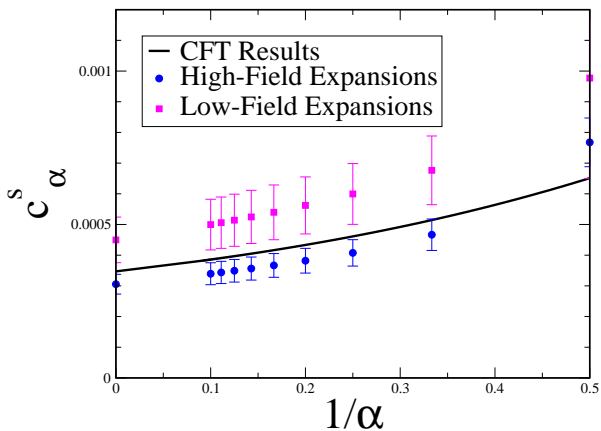}
\caption{\label{fig5} 
Log singularity coefficients in the corner entropy compared with $1/8$ of the log singularity coefficients for a sphere in
conformal field theory as a function of the Renyi index $\alpha$.
}
\end{center}
\end{figure}

We now turn to the key quantity of interest in the paper, namely, the corner entropy and its logarithmic singularity.
We expect that, on approach to the critical point, the corner entropy behaves as
\begin{equation}
c_\alpha = c_\alpha^s \ln{\xi}=-\nu \ c_\alpha^s \ln{(\lambda_c-\lambda)}.
\end{equation}
where $c_\alpha^s$ should be universal. By scaling, it should equal the coefficient $c_\alpha^s \ln{L}$ in the logarithmic
size dependence in an $L\times L \times L$ system, when the system is at the critical coupling\cite{max,nlc-dmrg}. We are interested in a
comparison with conformal field theory results for a free scalar-field in continuum\cite{cft1,huerta}.
The field theory result for one eighth of the corner logarithm, with a spherical interface,
is given by the expression\cite{cft1,huerta}
\begin{equation}
c_\alpha^s={1\over 720}{(1+\alpha)(1+\alpha^2)\over 4 \alpha^3}.
\end{equation}

To analyze the corner entropy, we need to take a derivative of the series, which converts a logarithmic
singularity into a simple pole. We then study this by a simple Pade approximant biasing the critical
point to the known value. First, consider the large $\alpha$ limit of our series. We estimate the corner singularity
coefficient $c_\infty^s$ to be $0.00031(4)$ and $0.00045(8)$ with $1/\lambda_c=5.15$ from high and low field sides respectively.
If we shift the critical point value to $1/\lambda_c=5.14$, we get $0.00032(4)$ and $0.00041(8)$ from the high and
low field sides respectively. These values are clearly consistent with the field theory value of ${1\over 2880}\approx 0.000347$.
A plot of the $\alpha$ dependence of $c_\alpha^s$, as obtained from our series analysis, is shown in Fig.~6.
The results bracket the field theory results from the two sides 
except for $\alpha=2$, where our analysis gives a somewhat larger value than the field theory\cite{footnote2}.
We find it quite remarkable that the results from a smooth spherical
boundary in a continuum theory closely approximates a corner calculation on a lattice\cite{tarun}.
In future, it would be interesting to study such corner entropies on other $3$-dimensional lattices.

In conclusion, in this paper we have calculated the entanglement entropies of the three-dimensional
transverse field Ising model by high and low field series expansions. 
Entropies associated with planes, edges
and corners in the boundary are separately calculated.
We have presented evidence that the series results are consistent with a transition at the
physical critical point from both sides.
We have also estimated the 
coefficients associated with the logarithmic singularity in the corner term.
They are quite close to $1/8$-th of the values obtained in conformal field theory for a spherical surface.
This suggests a universality between lattice and continuum models.
The correspondence between entanglement in lattice statistical models and continuum field theories deserves further
attention.

\begin{acknowledgements}
We would like to thank Anushya Chandran, Tarun Grover, Vedika Khemani, Matt Hastings, Roger Melko, Max Metlitski
and Shivaji Sondhi, for many enlightening communications.
This work is supported in part by NSF grant number  DMR-1004231 and DMR-1306048 and REU grant number PHY-1263201.
\end{acknowledgements}

%\bibliographystyle{apsrev}
%\bibliography{../bibinput/liter10}

\begin{thebibliography}{2}
\bibitem{footnote1} We should note that if the fictitious transition happens very close to
but not at the physical critical point, we may not be able to resolve it in our study.
\bibitem{footnote2} The differences for $\alpha=2$ could be due to numerical uncertainties from having a short series.
They could also arise from the fact that the Ising criticality in $3+1$ D is not completely equivalent to a free-field
theory at the upper critical dimension, and logarithmic corrections to critical properties are present.
\end{thebibliography}

\end{document}